\documentclass[12pt]{article}
\usepackage{amsthm,amsmath,amsfonts,amssymb}
\usepackage{graphicx}
\usepackage{enumerate}
\usepackage{natbib}
\usepackage{url} 

\usepackage{float}
\usepackage{commath,tikz,tabu}
\usepackage[utf8]{inputenc}
\usepackage{stackengine}
\usepackage{csquotes}

\usepackage{optidef}
\usepackage[procnumbered,ruled,vlined, noend]{algorithm2e}

\makeatletter
\AtBeginEnvironment{procedure}{\let\c@algocf\c@procedure}
\makeatother

\usepackage{amsmath,amssymb}
\usepackage[group-separator={,}]{siunitx}
\usepackage[flushleft]{threeparttable}
\usepackage{array,booktabs,makecell}

\newcommand{\R}{\mathbb{R}}
\newcommand{\D}{\mathcal{D}}

\newcommand{\one}{\mathbf{1}}

\newcommand{\betage}{\beta_{G \times E}}
\newcommand{\betagei}{\beta_{G_i\times E}}
\newcommand{\betagej}{\beta_{G_j\times E}}

\newcommand{\betag}{\beta_G}
\newcommand{\betagj}{\beta_{G_j}}
\newcommand{\betagi}{\beta_{G_i}}

\def\delequal{\mathrel{\ensurestackMath{\stackon[2pt]{=}{\text{def}}}}}

\begin{document}

\def\spacingset#1{\renewcommand{\baselinestretch}%
{#1}\small\normalsize} \spacingset{1}


  \title{\bf A scalable hierarchical lasso for gene-environment interactions}
  \author{Natalia Zemlianskaia,\\
    W. James Gauderman\\
    and \\
    Juan Pablo Lewinger \\
    Division of Biostatistics, Department of Preventive Medicine,\\
University or Southern California }
  \maketitle

\bigskip
\begin{abstract}
We describe a regularized regression model for the selection of gene-environment (G$\times$E) interactions. 
The model focuses on a single environmental exposure and induces a main-effect-before-interaction hierarchical structure. 
We propose an efficient fitting algorithm and screening rules that can discard large numbers of irrelevant predictors with high accuracy.  We present simulation results showing that the model outperforms existing joint selection methods for (G$\times$E) interactions in terms of selection performance, scalability and speed, and provide a real data application. Our implementation is available in the \texttt{gesso} R package.
\end{abstract}

\noindent%
\vfill

\newpage
\spacingset{1.5} 

\section{Introduction}

The problem of testing for the interaction between a single established predictor and a large number of candidate predictors arises in several contexts.  A common setting is the scanning for interactions in genomewide-association studies (GWAS), where the goal is to identify interactions between an established environmental risk factor (e.g. processed meat intake) and a large number of single nucleotide polymorphisms (SNPs), in relation to an outcome of interest (e.g. colorectal cancer) [\cite{gecco}]. Since many complex diseases have been linked to both genetic and environmental risk factors,  identifying gene-environment (G$\times$E) interactions, i.e. joint genetic and environmental effects beyond their component main effects,
is of great interest.  But the established predictor need not be an environmental risk factor and the candidate predictors can be other omic features like methylation and gene expression levels. For example, in a clinical trial setting, testing for the interaction between treatment and gene expression biomarkers can lead to the identification of subgroups with differential responses to a drug [\cite{gxe_treatment}].

In this paper we focus on the problem of scanning a large number of possible interactions with a fixed predictor. This is in contrast to the related but more challenging problem of exhaustive scanning all possible pairwise interactions within a set of predictors. Throughout the paper we will refer to the former as the gene-environment interaction selection problem, and we will use language specific to this context even though the proposed method is completely general and applies to any setting where the analytical goal is to identify interactions with a designated variable of special interest. Similarly, we will refer to the exhaustive pairwise testing problem as the gene-gene (G$\times$G) interaction selection problem. 

The dominant paradigm for genomewide interaction scans (GWIS) is to test each genetic marker one at a time for interaction with the risk factor of interest applying a stringent multiple testing correction to account for the number of tests performed. However, a joint analysis that simultaneously takes into account the effects of all markers is preferable to a one-at-a-time analysis where each marker is considered separately. Variables with weak effects might be more readily identifiable when the model has been adjusted for other causal predictors, and false positives may be reduced by the inclusion of stronger true causal predictors in the model [\cite{Ayers}]. Since GWAS data is high-dimensional (the number of SNPs is typically in the order of millions while the sample size is in the tens of thousands), a joint analysis of all markers using standard (unpenalized) multiple regression methods is not feasible. General regularized regression methods such as the Lasso [\cite{lasso_1996}] and Elastic Net [\cite{en_2005}] that are suitable for high-dimensional data and also perform variable selection can be used for G$\times$E identification but, because  they do not exploit the hierarchical structure of the (G$\times$E) problem, can perform suboptimally. A main-effect-before-interaction hierarchical structure ensures that the final selected model only includes interactions when  corresponding  main effects have been selected. This enhances the interpretability of the final model [\cite{Nelder}, \cite{Cox}] and increases 
the ability to detect intreactions by reducing the search space   [\cite{Chipman}].

Methods for the G$\times$G selection problem that exploit an interaction hierarchical structure like FAMILY [\cite{family}], glinternet [\cite{glinternet}], and hierNet [\cite{bien}], and their corresponding R implementations can be in principle applied to the G$\times$E case but they are optimized for the symmetric G$\times$G case, which results in vastly suboptimal performance in terms of run-time and scalability (they can handle at most a few hundred predictors) for the G$\times$E case. Indeed, the structure of the G$\times$E selection task is simpler because 1) the dimensionality of the problem grows linearly with additional environmental variables as opposed to quadratically for the G$\times$G selection task, and 2) because interactions with a single variable lead to a block-separable optimization problem, which, unlike the G$\times$G case, can be efficiently solved using a block coordinate descent algorithm. For these reasons, the G$\times$E selection problem is amenable to efficient implementations for large-scale analysis (e.g. genome-wide). However, efficient joint selection methods specific for G$\times$E in large-scale applications have not been developed. 

\cite{yale_2013} and \cite{yale_2017} adopted sparse group penalization approaches to accelerated failure time models for hierarchical selection of G$\times$E interactions, but their approaches do not scale to datasets with a very large number of predictors without additional pre-screening procedures. 


In this paper we present 
\texttt{gesso} (from G(by)E(la)sso) model for the hierarchical modeling of interaction terms.
We present an efficient fitting algorithm for the \texttt{gesso} model and powerful new screening rules that eliminate a large number of variables beforehand, making joint G$\times$E analyses feasible at genome-wide scale.

The paper is organized as follows. We first review the idea behind a hierarchical structure for interactions and present the \texttt{gesso} model. In section 3 we introduce screening rules and an adaptive convergence procedure we developed and incorporated into a block coordinate descent algorithm. We describe simulations in section 4 and a real data application in section 5 to demonstrate the applicability of the \texttt{gesso} model to large high-dimensional datasets and the scalability of our algorithm.

\section{Methods}
\subsection{Hierarchical structure}
The standard linear model for G$\times$E interactions with a single environmental exposure includes all interaction product terms between genetic variables and the environmental factor in addition to their marginal effects:
\begin{align} \label{standard_int}
E[Y] = \beta_0 + \beta_{E} E + \sum_{i=1}^{p} \beta_{G_i} G_i   +  \sum_{i=1}^{p} \beta_{G_i \times E}\ G_i \times E,
\end{align}
Here $Y \in \R ^n$ is a quantitative outcome of interest, $G$ is $n \times p$ matrix of genotypes, $G_i$ is a column of matrix $G$ corresponding to the $i$-th genotype, $E$ is the vector of environmental measurements of size $n$, 
$\beta_G \in \R ^p$ and
$\beta_E \in \R$ are the main effects, and
$\beta_{G \times E} \in \R ^{p}$ are the interaction effects.

A strong hierarchical structure implies that
if either the genetic or the environmental main effect is equal 
to zero, the corresponding interaction term has to be zero as well. In the G$\times$E context, the environmental predictor $E$ is usually chosen  because it has been previously identified as a risk factor. Thus, there is no need to maintain a hierarchical constraint with respect to the environmental effect, as it is known to have a main effect on the outcome. The strong hierarchical structure reduces then to
\begin{align*}
\beta_{G_i \times E} \ne 0 \Longrightarrow \beta_{G_i} \ne 0 \text{,  equivalently,  }
\beta_{G_i} = 0  \Longrightarrow \beta_{G_i \times E} = 0.   
\end{align*}

There are several ways to impose a hierarchical structure on the regression coefficients of model (\ref{standard_int}). One approach is forward selection [\cite{FSTEP_1}], 
a procedure that iteratively considers adding the \textquote{best} variable to the model ensuring that an interaction term can be added only if its main effects have already been added to the model on a previous iteration. Since step-wise selection is a  greedy algorithm, it tends to underperform compared to global optimization approaches [\cite{Ayers}]. Another way is to reparameterize the interaction coefficients as $\beta_{G_i \times E} = \gamma_{i} \beta_{G_i} \beta_{E}$, where $\gamma_{i}$ is an introduced model parameter [\cite{sail}, \cite{yale_2020}], but this results in a non-convex objective function. Regularization  can also be used to impose a desired hierarchical structure [\cite{cap}] and this is the approach we follow. 

\subsection{The \texttt{gesso} model}
Denote the mean square error loss function for a G$\times$E problem by $$q(\beta_0, \beta_G, \beta_E, \beta_{G\times E}) = \frac{1}{2n} \norm{Y - \big( \beta_0 + \sum_{i} \beta_{G_i} G_i +  \beta_{E} E +  \sum_i \beta_{G_i \times E} \ G_i \times E \big)}_2^2.$$ 
We are interested in hierarchical selection of the $G_i \times E$ interaction terms  that are associated with the outcome $Y$. We propose the following model that we call \texttt{gesso} (G(by)E(la)sso):
\begin{align} \label{unconstrained}
\mathrel{\ensurestackMath{\stackunder[1pt]{\text{minimize}}{{\beta_0, \beta_G, \beta_E, \beta_{G\times E}}}}}
q(\beta_0, \beta_G, \beta_E, \beta_{G\times E}) +  \sum_{i=1}^p  \Big ( \lambda_1 \norm{(\beta_{G_i}, \beta_{G_i \times E})}_{\infty}   +  \lambda_2 |\beta_{G_i \times E}| \Big ).
\end{align}

The model has several important properties. First, it guarantees the desired hierarchical relationships between genetic main effects and interaction effects, since the penalty satisfies the overlapping group hierarchical principle [\cite{cap}]. Having $\beta_{G_j\times E}$ in every group where $\beta_{G_j}$ is present ensures that once $\beta_{G_j\times E}$ deviates from zero, the penalty for $\beta_{G_j}$ becomes close to zero by the properties of the group lasso models. 
In addition, having $\beta_{G_j\times E}$ in a group of its own makes it possible for $\beta_{G_j}$ to deviate from zero, when $\betagej$ is zero. Second, the regularized objective function is convex, which can take advantage of 
convex optimization theory and 
algorithms that ensure convergence to a global optimal solution. Moreover, the block-separable structure of the problem, with small-block sizes allows for highly efficient solvers. Third, we include two tuning parameters in the model, $\lambda_1$ and $\lambda_2$ enabling flexible and decoupled data dependent control over the group and interaction penalties.  
Lastly, the group $L_{\infty}$ penalty has a connection to a Lasso model with hierarchical constraints that we discuss next.

\subsection{Connection to the hierarchical Lasso}
\cite{bien} proposed a 
Lasso model with hierarchical constraints for all pairwise interactions (the G$\times$G selection problem) and demonstrated its advantages over the standard Lasso. 
The authors showed that the hierarchical Lasso model is equivalent to the unconstrained overlapping group lasso model with an $L_{\infty}$ group norm. Following Bien et al. it can be shown that our proposed model (\ref{unconstrained}) is equivalent to a constrained model (\ref{relaxation}). We describe  
the intuition behind the constrained version of the model (\ref{relaxation}) below and provide a proof of equivalence of the two models in the Appendix A. 
\subsubsection*{Lasso with G$\times$E Hierarchical Constraints}
To impose the desired hierarchical structure, the constraints $|\beta_{G_i \times E}| \le |\beta_{G_i}|$ can be added to the standard Lasso model. These ensure the main-effect before interaction property:\\  $\beta_{G_i} = 0 \Longrightarrow \beta_{G_i \times E} = 0$. Furthermore, the model makes the implicit (and arguably a reasonable) assumption that important interactions have large main effects.  When this assumption is met, the model will be more powerful in detecting interactions. Unfortunately, the constraint set above is non-convex and yields the non-convex problem: 
\begin{mini}|l|
{\beta_0, \beta_G, \beta_E, \beta_{G\times E}}{q(\beta_0, \beta_G, \beta_E, \beta_{G\times E}) + \lambda_1 \norm{\beta_{G}}_1 + \lambda_2 \norm{\beta_{G\times E}}_1}
{\label{hiernet_gxe}}{}
\addConstraint{|\beta_{G_i \times E}| \le |\beta_{G_i}|, \text{ for } i = 1, ..., p}.
\end{mini}

In order to transform the non-convex optimization problem (\ref{hiernet_gxe}) into a convex one, we decompose $\beta_{G} $ as $\beta_{G}=\beta_{G}^+ - \beta_{G}^- $ and  $|\beta_{G}| $ as $\beta_{G}^+ + \beta_{G}^-$, where $\beta_{G}^+ \ge 0$ and $\beta_{G}^- \ge 0$. The  non-convex constraints $|\beta_{G_i \times E}| \le |\beta_{G_i}|$ are then replaced by the convex constraints $|\beta_{G_i \times E}| \le \beta_{G_i}^+ + \beta_{G_i}^-$. Note that
$|\beta_{G}| = \beta_{G}^+ + \beta_{G}^-$ only if $\beta_G^+ \beta_G^- \equiv 0$, so removing the latter condition results in a relaxed formulation that is not equivalent to the original non-convex problem.
Substituting $\beta_{G} = \beta_{G}^+ - \beta_{G}^- $ and $|\beta_{G}|$ to $\beta_{G}^+ + \beta_{G}^- $ in (\ref{hiernet_gxe}) we obtain a convex relaxation of the model:
\begin{mini}|l|
{\beta_0, \beta_G^+, \beta_G^-, \beta_E, \beta_{G\times E}}{q(\beta_0, \beta_G^+, \beta_G^-, \beta_E, \beta_{G\times E}) + \lambda_1 \one^T ( \beta_{G}^+ + \beta_{G}^- ) + \lambda_2 \norm{\beta_{G\times E}}_1}
{\label{relaxation}}{}
\addConstraint{|\beta_{G_i \times E}| \le \beta_{G_i}^+ + \beta_{G_i}^-, \ \beta_{G_i}^+ \ge 0, \  \beta_{G_i}^- \ge 0\text{ for } i = 1, ..., p.}
\end{mini}
Now the constraint set is convex and the optimization problem is convex as well. 
The relaxed constraints ($\betagi^+ + \betagi^- \ge |\betagei|$) are less restrictive than the original $|\betagi| \ge |\betagei|$. In particular, the model can yield a large   $|\betagei|$ estimate and a moderately sized $|\betagi|$ by making both $\betagi^+$ and $\betagi^-$ large. 

Examining the equivalent formulations for \texttt{gesso} (unconstrained group $L_{\infty}$ model (\ref{unconstrained}) and constrained model (\ref{relaxation})) we can see that the group $L_{\infty}$ norm corresponds to the constraints on the effect sizes of the interactions and the main effects. Intuitively,
 because of the connection between the relaxed model (\ref{relaxation}) and model (\ref{hiernet_gxe}),  (\ref{relaxation}) will be more powerful for detecting interactions 
 in the case when important interactions have large main effects.
This is confirmed by our simulation further below. 

Additionally, the constrained formulation of \texttt{gesso} (\ref{relaxation}) allows for a simpler, interpretable form for the coordinate-wise solutions, the dual problem, and the development of the screening rules.

\section{Block coordinate descent algorithm for \texttt{gesso}}

\cite{friedman_cd_1} have proposed using cyclic coordinate descent for solving convex regularized regression problems involving $L_1$ and $L_2$ penalties and their combinations. The coordinate descent algorithm is particularly advantageous when each iteration  involves  only fast analytic updates. In addition, screening rules that exploit the sparse structure induced by the penalties can be readily incorporated into the algorithm to eliminate a large number of variables beforehand, making it much faster than alternative convex optimization algorithms.

For convex block-separable functions, convergence of the coordinate descent algorithm to a global minimum is guaranteed [\cite{bcd_convergence}]. The \texttt{gesso} model has a convex objective function with a smooth loss component and a non-smooth separable penalty component, where each block consists of $\betagj$ and  $\betagej$.  Thus, the model can be fitted using a block coordinate descent (BCD) algorithm and convergence to a global optimal solution is guaranteed. Briefly, BCD  optimizes the objective by cycling through the coordinate blocks $1, ..., p$ and minimizing the objective along each coordinate block direction while keeping all other blocks fixed at their most current values. The coordinate-wise updates for model (\ref{relaxation}) can be obtained by working with the Lagrangian version of the model. The derivations are provided in the supplementary materials (section 1). The rest of the section focuses on new efficient screening rules we developed specifically for \texttt{gesso}.

\subsection{Dual formulation of \texttt{gesso}}


Consider the following primal formulation of the \texttt{gesso} problem obtained by substituting the residuals for a new variable $z$:
\begin{mini}|l| 
{\beta^{+}, \beta^{-}, \betage, z }{\frac{1}{2n} z^T z + \lambda_1 \one^T(\beta^+ + \beta^-) + \lambda_2 \Arrowvert \betage \Arrowvert_{1}}
{\label{primal}}{}
\addConstraint{| \betage |
\preccurlyeq (\beta^+ + \beta^-), \ \beta^{+} \succcurlyeq 0, \ \beta^{-} \succcurlyeq 0 }
\addConstraint{z = Y - \Big(\one \beta_0 + G (\beta^+ - \beta^-) + E\beta_E + (G\times E) \betage\Big),}
\end{mini}
The $\succcurlyeq$ symbol denotes an element-wise comparison ($x \succcurlyeq 0 \iff  x_j \ge 0, \text{ for } j = 1,... , p$), $G\times E$ is a column-wise matrix of interaction vectors $G_i \times E$. In order to formulate the dual problem we introduce the dual variables $ \delta \in \R^p, \ \delta \succcurlyeq 0$, associated with the constraint $|\betage| \preccurlyeq (\beta^+ + \beta^-)$,
and
$\nu \in \R^n$, associated with the constraint $z = Y - \Big(\one \beta_0 + G (\beta^+ - \beta^-) + E\beta_E + G\times E \betage\Big)$. Substituting the residuals for a new variable in the primal formulation is a common approach for deriving a dual formulation.
For simplicity, we denote the linear predictor $\one \beta_0 + G (\beta^+ - \beta^-) + E\beta_E + (G\times E) \betage$ by $X\beta$. Based on the alternative primal formulation the dual takes the simple form below (section 2 of the supplementary materials):
\begin{maxi*}|l|
{\delta, \nu}{\frac{n}{2} \left( \norm{\frac{Y}{n}}_2^2 - \norm{\frac{Y}{n} - \nu}_2^2  \right)}
{}{}
\addConstraint{\delta, \nu \in \D_F,}
\end{maxi*}
where we denote the dual feasible region as
$
\D_F :=
    \begin{cases} 
    |\nu^T G\times E| \preccurlyeq \lambda_2 + \delta, \\
|\nu^T G| \preccurlyeq \lambda_1 - \delta,\\
\delta  \in [0, \lambda_1], 
    \end{cases}  
$\\
and the dual objective as $D(\nu) =\frac{n}{2} \left( \norm{\frac{Y}{n}}_2^2 - \norm{\frac{Y}{n} - \nu}_2^2  \right).$

It follows that the optimal solution of the dual problem $\hat{\nu}$ can be viewed as a projection of $\frac{Y}{n}$ onto the dual feasible set $\D_F$:
$$
\hat{\nu} = \arg\max_{\nu \in \D_F}\ \frac{n}{2} \left( \norm{\frac{Y}{n}}_2^2 - \norm{\frac{Y}{n} - \nu}_2^2  \right) = \arg\min_{\nu \in \D_F}\ \norm{\frac{Y}{n} - \nu}_2^2 \delequal \mathrm{Proj}_{\D_F}\left( \frac{Y}{n} \right).
$$
And from the stationarity conditions we establish the following important relationship:
\begin{align} \label{primal-dual}
 \hat{\nu} = \frac{\hat{z}}{n} = \frac{Y - X\hat{\beta}}{n}.
\end{align}
The optimal dual variable $\hat{\nu}$ equals the residuals scaled by the number of observations. Equation (\ref{primal-dual}) defines the link between the primal ($\hat{\beta}$) and the dual ($\hat{\nu}$) optimal solutions. The stationarity and complementary slackness conditions for $\betagi$ and $\betagei$ (section 2 of supplementary materials) lead to the following important consequences for the primal and dual optimal variables:
\begin{align} 
\Big|\hat{\nu} ^T (G_i\times E) \Big| < \lambda_2 + \hat{\delta}_i \implies \hat{\beta}_{G_i\times E} = 0 \text{, for all $i$,} \label{beta_gxej_zero}\\
\Big|\hat{\nu} ^T G_i \Big| < \lambda_1 - \hat{\delta}_i \implies \hat{\beta}_{G_i} = 0 \text{, for all $i$.} \label{beta_j_zero}
\end{align}
Conditions (\ref{beta_gxej_zero}), (\ref{beta_j_zero})
form the basis for the screening rules we develop later in this section. Specifically, we exploit the above conditions to identify null predictors and avoid spending time cycling through them in the BCD algorithm. This leads to a substantial computational speed-up, especially for high-dimensional sparse problems.


\subsection{Screening rules}
Screening rules are used to identify predictors that are strongly associated with the outcome and screen out those that are likely to be null. Incorporation of the screening rules to the coordinate descent algorithm can greatly improve computational speed, making large but sparse high-dimensional problems computationally tractable. In this section we first describe the SAFE screening rules for \texttt{gesso} following the principles of the Lasso SAFE rules [\cite{safe}], which form the stepping stone for the more efficient screening rules we developed and focus on later. 

\subsubsection{SAFE rules for \texttt{gesso}}
\cite{safe} proposed the SAFE screening rules for the Lasso model that guarantee a coefficient will be zero in the solution vector. In this section we derive SAFE rules to screen predictors for the \texttt{gesso} model.  

By the KKT conditions (\ref{beta_gxej_zero}) and (\ref{beta_j_zero})  
\begin{align} \label{safe_start1}
\begin{rcases} 
\Big|\hat{\nu} ^T G_i \Big| < \lambda_1 - \delta_i \\
\Big|\hat{\nu} ^T (G_i\times E) \Big| < \lambda_2 + \delta_i\\
\delta_i \in [0, \lambda_1]
\end{rcases}
\implies \hat{\beta}_i = \hat{\beta}_{G_i \times E} = 0
\end{align}
for any primal optimal variables $\hat{\beta}_{G_i}$ and $\hat{\beta}_{G_i\times E}$, dual optimal variable $\hat{\nu}$, and dual feasible variable $\delta_i$ (for a fixed optimal $\hat{\nu}$ any feasible $\delta_i$ would also be optimal, since the dual objective function does not depend on $\delta$). The problem is that we do not know the dual optimal variable $\hat{\nu}$. 

The idea is to
create upper bounds for $|\hat{\nu}^T G_i|$ and $|\hat{\nu}^T (G_i\times E)|$ that are easy to compute. In particular, 
consider a dual feasible variable $\nu_0$ and  denote $D(\nu_0)$ as $\gamma $.
Let $\Theta = \{\nu : \ D(\nu) \ge \gamma\}$. As $\hat{\nu} = \text{argmax}_{\D_F} D(\nu)$ we have $D(\hat{\nu}) \ge D(\nu_0) = \gamma$ and, hence,
$\hat{\nu} \in \Theta$. Then  
$
\max_{\nu \in \Theta} |\nu ^T G_i| < \lambda_1 - \delta_i \implies |\hat{\nu} ^T G_i| < \lambda_1 - \delta_i.
$ 
The same is true for the interaction terms 
$
\max_{\nu \in \Theta} |\nu ^T (G_i\times E)| < \lambda_2 + \delta_i \implies |\hat{\nu} ^T (G_i\times E)| < \lambda_2 + \delta_i,
$
where $\max_{\nu \in \Theta} |\nu ^T G_i|$ and $\max_{\nu \in \Theta} |\nu ^T (G_i\times E)|$ are the desired upper bounds for $|\hat{\nu}^T G_i|$ and $|\hat{\nu}^T (G_i\times E)|$ respectively.
The above arguments lead to the following rules: 
\begin{align} 
&\max_{\nu \in \Theta} |\nu ^T G_i| < \lambda_1 - \delta_i \implies \hat{\beta}_{G_i} = 0, \label{safe_g}\\
&\max_{\nu \in \Theta} |\nu ^T (G_i\times E)| < \lambda_2 + \delta_i \implies \hat{\beta}_{G_i \times E} = 0.\label{safe_gxe}
\end{align}
Note that the region $\Theta = \{\nu : \ D(\nu) \ge \gamma\}$ is equivalent to  
$\frac{n}{2} \left( \norm{\frac{Y}{n}}^2_2 -\norm{\frac{Y}{n} - \nu}^2_2\right) \ge \gamma \iff$
$\norm{\frac{Y}{n} - \nu}^2 \le \norm{\frac{Y}{n}}^2 - \frac{2}{n}\gamma
$ which is the equation of an Euclidean ball with $r^2 = \norm{\frac{Y}{n}}^2 - \frac{2}{n}\gamma$ and centered at $\frac{Y}{n}$. Note that
$
r = \sqrt{\norm{\frac{Y}{n}}^2 - \frac{2}{n}\gamma} = \sqrt{\norm{\frac{Y}{n}}^2  - \frac{2}{n} \frac{n}{2} \left( \norm{\frac{Y}{n}}^2_2 + \norm{\frac{Y}{n} - \nu_0}^2_2\right)
} = \norm{\frac{Y}{n} - \nu_0} .
$
We denote this ball as $B(c, r)$, where $c = \frac{Y}{n}$ and $r = \norm{\frac{Y}{n} - \nu_0}$.
As a consequence, the upper bounds we constructed $\max_{\nu \in \Theta} |\nu ^T G_i|$ and $\max_{\nu \in \Theta} |\nu ^T (G_i\times E)|$ are equivalent to the following optimization problems:
$\max_{\nu \in B(c, r)} |\nu ^T G_i|$ and
$\max_{\nu \in B(c, r)}  |\nu ^T (G_i\times E)|$ 
and the region $\Theta$ is the ball $B(c, r)$, which has two important properties:

(1) it contains the optimal dual solution $\hat{\nu} \in \Theta$,

(2) it results in closed form solutions for the desired upper bounds (\ref{safe_g}) and (\ref{safe_gxe}).\\ Specifically, 
for a general $X \in \R^n$ the solution  to $\max_{\nu \in B(c , r)} |\nu^T X|$  is 
$
\max_{\nu \in B(c , r)} |\nu^T X| = r  \norm{X} + |X^T c|.
$
Leaving the derivations to Appendix C, the SAFE rules to discard $(\betagi, \beta_{G_i \times E})$ are given by:
\begin{align} \label{SAFE_g_gxe}
\max \big\{0, \ r  \norm{G_i\times E} + |(G_i\times E)^T c| -  \lambda_2\big\} < \lambda_1 - r  \norm{G_i} - |G_i^T c|
\implies \hat{\beta}_{G_i} = \hat{\beta}_{G_i \times E} = 0.
\end{align}

To complete the construction of the SAFE rules, we need to find a dual feasible point $\nu_0 \in \D_F$ to calculate $r = r(\nu_0)$. 
We can naturally obtain a dual point $\nu_0$ given the current estimate $\beta$ via the primal-dual link from the stationarity conditions (\ref{primal-dual}). Denote $\nu_{\mathrm{res}}(\beta) = \frac{Y - X\beta}{n}$, where $res$ stands for residuals. In general, $\nu_{\mathrm{res}}(\beta)$ will not necessary be feasible, thus, we re-scale $\nu_{\mathrm{res}}(\beta)$ to ensure it is in the feasible region $\D_F$. For example, for $\delta = 0$ we can consider the re-scaling factor $x = \min \Big(\frac{\lambda_1}{|\nu_{\mathrm{res}}(\beta)^T G_i|}, \frac{\lambda_2}{|\nu_{\mathrm{res}}(\beta)^T G_i \times E|} \Big)$ so that $\nu_0 = x \nu_{\mathrm{res}}(\beta)$ is feasible. We call the value $x \nu_{\mathrm{res}}(\beta)$ a \textit{naive} projection of $\nu_{\mathrm{res}}$. In the next section we show an alternative way to re-scale $\nu_{\mathrm{res}}(\beta)$  to obtain a better feasible point.

\subsubsection{Optimal naive projection} \label{section_naive}
The better we choose our feasible $\nu_0$ and $\delta$ (i.e. closer to the optimal solution) the tighter our upper bounds for $|\hat{\nu}^T G_i|$ and $|\hat{\nu}^T (G_i\times E)|$ on the set  $\Theta$ = $\{\nu: D(\nu) \ge D(\nu_0)\}$ are. In the previous re-scaling of $\nu_{\mathrm{res}}(\beta)$ we set the dual variable $\delta$ equal to 0 for simplicity when we constructed a dual feasible point $\nu_0$. However, we can find a point $\nu_0$ closer to the optimal by changing $\delta$ to our advantage. 

Formally, we want to find a scalar $x$ such that $\nu_0 = x \nu_{\mathrm{res}}(\beta)$ is feasible and which minimizes $D(\nu_0)$. 
This leads to the  following optimization problem with respect to $x$ and $\delta$:
\begin{mini}|l| 
{x, \delta}{ D(\nu_0)}
{\label{safe_new}}{}
\addConstraint{|\nu_0^T G\times E| \preccurlyeq\lambda_2 + \delta}
\addConstraint{|\nu_0^T G| \preccurlyeq \lambda_1 - \delta}
\addConstraint{\delta  \succcurlyeq 0}
\addConstraint{\text{where } \nu_0 = x \nu_{\mathrm{res}}(\beta).}
\end{mini}

We present the closed-form solution to the optimization problem (\ref{safe_new}) in Appendix B, which we
use to obtain a dual feasible variable $\nu_0$ ($\nu_0 = x \nu_{\mathrm{res}}(\beta)$) that we require for the SAFE rules (\ref{SAFE_g_gxe}).

\subsubsection{Warm starts and dynamic screening}
The penalty parameters $\lambda_1$ and  $\lambda_2$ will be typically tuned by cross-validation. In practice, this requires  fitting  the model for a grid of $\lambda_1$ and $\lambda_2$ values. A common choice is a logarithmic grid of 30 to 100 consecutive points. Maximum grid value can be determined from the stationarity KKT conditions.
When solutions are computed along such a sequence or path of tuning parameters,
warm starts is a standard approach to reduce the number of coordinate descent iterations required to achieve convergence.
Warm starting refers to using the previously computed solution $\hat{\beta}(\lambda^{(k-1)})$ with respect to a grid of tuning parameters to initialize the parameters values for the coordinate descent algorithm $\beta^{(k)}$ on the next step. 




\begin{figure}[ht!] 
\centering
\includegraphics[width=1\textwidth]{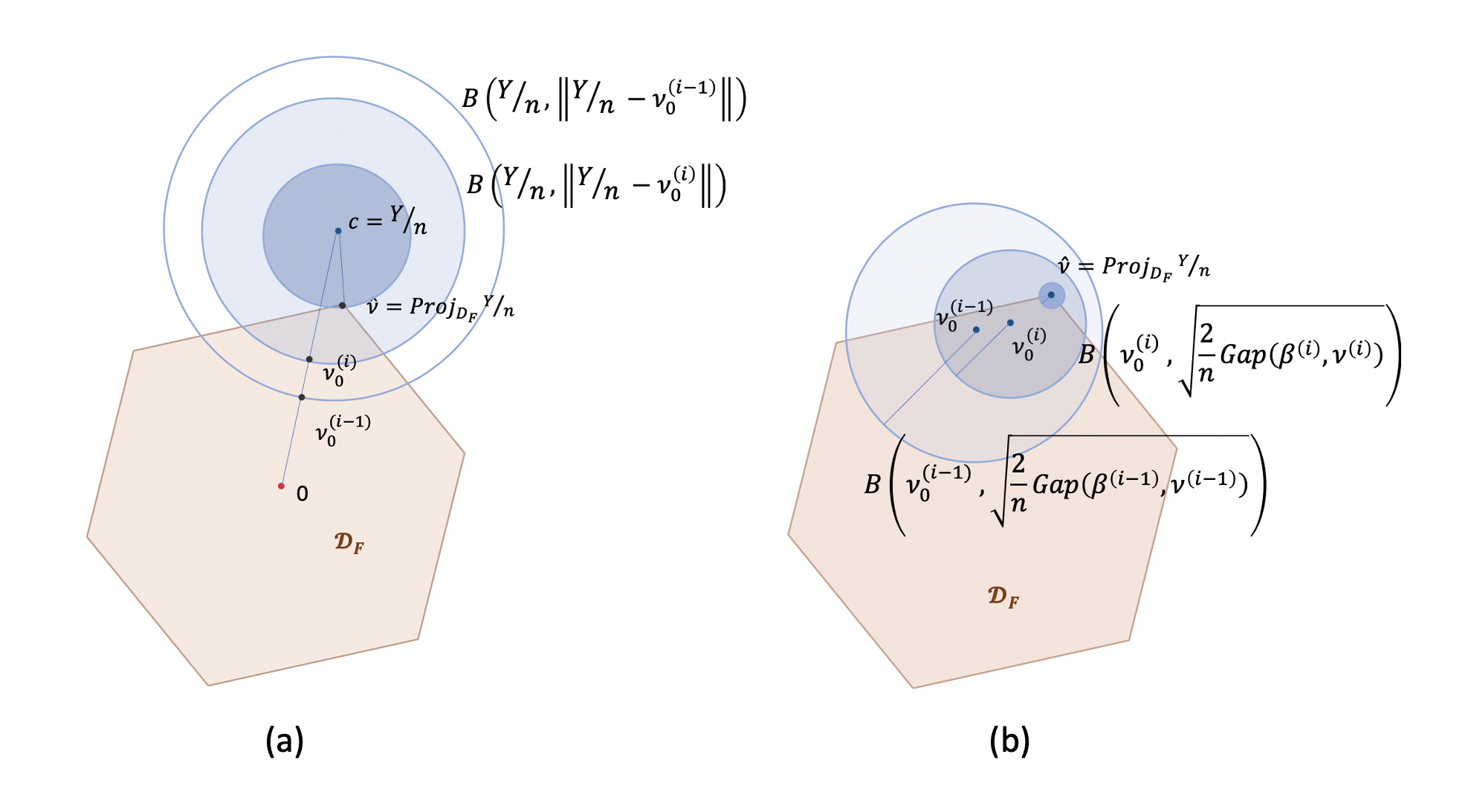}
\caption{(a) Dynamic SAFE regions, (b) dynamic GAP SAFE regions.}
\label{dynamic_safe}
\end{figure}


The idea behind dynamic screening [\cite{dynamic}] is to iteratively improve the dual feasible point $\nu_0$ 
during the coordinate descent updates.
The residuals $Y - X\beta$ are updated at each iteration $i$ of the coordinate descent algorithm, and since the coordinate descent algorithm guarantees convergence, we have that
\begin{align} \label{conv}
\beta^{(i)}  \underset{i}{\longrightarrow} \hat{\beta} \text{  and  } \nu_{\mathrm{res}}^{(i)} = \frac{Y - X\beta^{(i)}}{n} \underset{i}{\longrightarrow} \frac{Y - X\hat{\beta}}{n} = \hat{\nu} \implies \nu_{\mathrm{res}}^{(i)} \underset{i}{\longrightarrow} \hat{\nu}, \ \ \nu_{0}^{(i)} \underset{i}{\longrightarrow} \hat{\nu}.
\end{align}

Recall that our SAFE rules depend on the feasible point $\nu_0$ through the radius of the spherical region on which we base the upper bounds : $B(c, r(\nu_0^{(i)})) = B\Big(\frac{Y}{n}, \norm{\frac{Y}{n} - \nu_0^{(i)}}\Big)$. Also, recall that $\hat{\nu} = \mathrm{Proj}_{\D_F} \frac{Y}{n} = \min_{\nu \in \D_F} \norm{\frac{Y}{n} - \nu}^2,$ $\hat{\nu}$ is the closest feasible point to the center $\frac{Y}{n}$.
Hence, having $\nu_0$ closer to the optimal $\hat{\nu}$ reduces the radius of the sphere and ensures tighter upper bounds $\max_{\nu \in B(c, r)} |\nu ^T G_i|$ and
$\max_{\nu \in B(c, r)}  |\nu^T (G_i\times E)|$ for $|\hat{\nu} ^T G_i|$ and
$|\hat{\nu} ^T (G_i\times E)|$. This, in turn, ensures better SAFE rules that are able to discard more variables. By performing the
SAFE rules screening not only at the beginning of each new iteration $\lambda_k = (\lambda_1, \lambda_2)_k$, but all along the iterations $i$ for the coordinate descent, we iteratively improve the estimate of $\hat{\nu}$ with $\nu_0^{(i)}$ and consequentially keep improving our SAFE rules. 

The proposed procedure is the following: at each iteration $i$ of the algorithm use the current residuals $Y - X\beta^{(i)}$ to obtain a current estimate of the dual variable $\nu_{\mathrm{res}}^{(i)} = \frac{Y - X\beta^{(i)}}{n}$, naively project it onto the feasible set $\D_F$ and apply the SAFE rules (\ref{SAFE_g_gxe}). Figure \ref{dynamic_safe}(a) illustrates the iterative process of constructing the SAFE spherical regions. One concern is how expensive it is to compute the naive projection and the SAFE rules at each iteration, which we address in details in section \ref{section_max_diff}.

However, there are clear disadvantages of our current choice of center $c = \frac{Y}{n}$ and radius $r = \norm{\frac{Y}{n} - \nu_0^{(i)}}$. Even when the dual feasible variable $\nu_0^{(i)}$ converges to the optimal value $\hat{\nu}$, the ball $B \big(\frac{Y}{n}, \norm{\frac{Y}{n} - \hat{\nu}}\big)$ does not shrink around the dual optimal variable, since the radius does not converge to zero and the center is static, indicating that our upper bounds  $\max_{\nu \in B \big(\frac{Y}{n}, \norm{\frac{Y}{n} - \hat{\nu}}\big)}|\nu ^T G_i|$ and $\max_{\nu \in B \big(\frac{Y}{n}, \norm{\frac{Y}{n} - \hat{\nu}}\big)}|\nu ^T (G_i\times E)|$ remain loose (Figure \ref{dynamic_safe}(a)). The Gap SAFE rules proposed by \cite{gap_original} beautifully address the above disadvantages. We present the ideas behind the Gap SAFE rules and their application to the \texttt{gesso} model in the next section.

\subsubsection{Gap SAFE rules for \texttt{gesso}}

Denote the primal objective function for the \texttt{gesso} model as $P(\beta)$:
\begin{align} \label{primal_objective}
P(\beta) = \frac{1}{2n}\norm{Y - X\beta}^2  + \lambda_1 \sum_{i=1}^p \max \big\{ |\beta_{G_i}|, |\betagei|_1  \big\}  + \lambda_2 \norm{\beta_{G \times E}}_1
\end{align} and the dual objective as $D(\nu)$:
\begin{align} \label{dual_objective}
D(\nu) = \frac{n}{2} \Bigg( \norm{\frac{Y}{n}}^2_2 -\norm{\frac{Y}{n} - \nu}^2_2 \Bigg).
\end{align}
The duality gap, denoted as $\mathrm{Gap}(\beta, \nu)$ is the difference between the primal and the dual objectives
$\mathrm{Gap}(\beta, \nu)=  P(\beta) - D(\nu). $
For the optimal solutions we have $P(\beta) \ge P(\hat{\beta})$ and $D(\nu) \le D(\hat{\nu})$.
By weak duality 
$
D(\nu) \le D(\hat{\nu}) \le P(\hat{\beta}) \le P(\beta) \implies \mathrm{Gap}(\beta, \nu) \ge 0.
$
The duality gap provides an upper bound for the suboptimality gap $P(\beta) - P(\hat{\beta})$ as 
$P(\beta) - P(\hat{\beta}) \le P(\beta) - D(\nu) = \mathrm{Gap}(\beta, \nu) \le \epsilon$.
 Therefore, given a tolerance $\epsilon > 0$, if at iteration $t$ of the BCD
algorithm we can construct $\beta^{(t)}$ and $\nu^{(t)} \in \D_F$ such that 
$\mathrm{Gap}(\beta^{(t)}, \nu^{(t)}) \le \epsilon$, then $\beta^{(t)}$ is guaranteed to be a $\epsilon$-optimal solution of the primal problem.
Note that in order to use the duality gap at iteration $t$ as a stopping criterion we need a dual feasible point $\nu^{(t)} \in \D_F$. We, again, utilize the naive projection method and obtain $\nu_0^{(t)}$.



Because $D(\nu)$ is a quadratic and strongly concave function with concavity modulus $n$ [\cite{convex_optim}] we have:
$
D(\nu) \le D(\hat{\nu}) \ + \langle \nabla D(\hat{\nu}), \nu - \hat{\nu} \rangle - \ \frac{n}{2} \norm{\nu - \hat{\nu}}^2, 
$
$$
\frac{n}{2} \norm{\nu - \hat{\nu}}^2 \le
D(\hat{\nu}) - D(\nu) \ +  \langle \nabla D(\hat{\nu}), \nu - \hat{\nu} \rangle \ \le 
P(\beta) - D(\nu) = \mathrm{Gap}(\beta, \nu),
$$
where the second inequality follows from weak duality and the optimality conditions for $\hat{\nu}$. Thus, 
\begin{align} \label{dual_r}
 \norm{\nu - \hat{\nu}} \le \sqrt{\frac{2}{n}\mathrm{Gap}(\beta, \nu)}.
\end{align}

Gap SAFE rules work with the Euclidean ball with center $c = \nu^{(i)}_0$ and radius $r = \sqrt{\frac{2}{n}\mathrm{Gap}(\beta^{(i)}, \nu^{(i)}_0)}$, which we denote as $B_{\mathrm{Gap}}(c, r)$. This is a valid region since $\hat{\nu} \in B_{\mathrm{Gap}}(c, r)$ by (\ref{dual_r}). An important consequence of the above construction is that when $\beta^{(i)} \underset{i}{\longrightarrow} \hat{\beta}$, then $\nu_0^{(i)} \underset{i}{\longrightarrow} \hat{\nu}$  via the primal-dual link (\ref{primal-dual}, \ref{conv}) and $\mathrm{Gap}(\beta^{(i)},\nu_0^{(i)}) \underset{i}{\longrightarrow} 0$.
Figure \ref{dynamic_safe}(b) illustrates that the dynamic approach discussed in the previous section in combination with the Gap SAFE rules very naturally results in improving the upper bounds $\max_{B_\mathrm{Gap}(c, r)}|\nu ^T G_i|$ and $\max_{B_\mathrm{Gap}(c, r)}|\nu ^T (G_i\times E)|$, since the radius of our new Gap ball converges to zero and the center converges to the dual optimal point. The Gap SAFE rules follow from substituting $r = \sqrt{\frac{2}{n}\mathrm{Gap}(\beta^{(i)}, \nu^{(i)}_0)}$ and $c=\nu^{(i)}_0$ in (\ref{SAFE_g_gxe}).

\subsubsection{Working set strategy}
\cite{celer1} proposed to use a working set approach with the Gap SAFE rules to achieve substantial speedups over the state of art Lasso solvers, including glmnet. 
The working set strategy involves two nested iteration
loops. In the outer loop, a set of predictors $W_t \subset \{ 1,.., p \}$ is defined, called a working set (WS). In the inner loop, the coordinate descent algorithm is launched to solve
the problem restricted to $X_{W_t}$ (i.e. considering only the predictors in $W_t$).

We adopt the Gap SAFE rules we developed for the \texttt{gesso} model to incorporate the proposed working set strategy. Recall the SAFE rules we constructed in (\ref{SAFE_g_gxe}):
$$
\max \big\{0, \ r  \norm{G_i\times E} + |(G_i\times E)^T c| -  \lambda_2\big\} < \lambda_1 - r  \norm{G_i} - |G_i^T c| \implies \hat{\beta}_{G_i} = \hat{\beta}_{G_i \times E} = 0.
$$
By simply rearranging the terms in the inequality we get:
$$
\frac{\lambda_1  - |G_i^T c| + \max \big\{r  \norm{G_i\times E}, \lambda_2 - |(G_i\times E)^T c| \big\}}{\norm{G_i} +   \norm{G_i\times E}} > r. 
$$
Define the left-hand side of the above inequality as $d_i$. 
Note that our SAFE rules (\ref{SAFE_g_gxe}) are equivalent to $d_i > r$. 

The idea is that $d_i$ now represents a score for how likely the predictor is zero or non-zero based on the Gap SAFE rules.
Predictors for which $d_i$ is small are more likely to be non-zero and, conversely, predictors with larger values of $d_i$ are more likely to be zero up to the point when $d_i > r$ and the corresponding predictor is exactly zero by the SAFE rules. Here by predictor we mean the pair $(\beta_{G_i}, \betagei)$.

The proposed procedure is as follows: compute the initial number of variables to be assigned to the working set ($working\_set\_size$). Calculate $d_i$ and define the working set as the indices of the smallest $d_i$ values up to the value of $working\_set\_size$. Fit the coordinate descent algorithm on the variables from the working set only and check if we achieved an optimal solution via the duality gap for all of the variables. If not, increase the size of the working set, recalculate $d_i$ according to the new estimates obtained by fitting on a previous working set, and repeat the procedure. We increase the size of the working set by two each time.


Algorithm 1 presents the block coordinate descent algorithm and the stopping criterion we propose to use. Algorithm 2 describes the main steps of the working set strategy. To recapitulate, the main advantage of the algorithm is that we select variables that are likely to be non-zero and leave likely zeroes out so we do not have to spend unnecessary  time fitting them. This approach can be thought of as an acceleration of the Gap SAFE rules.

\vskip0.1in
\begin{footnotesize}
\setcounter{algocf}{0}
\begin{algorithm}[H]
\SetAlgoLined
 \For{i = 1, ... , max\_iter}{
    check\_duality\_gap($\beta$) \textcolor{blue}{//convergence criterion, computationally expensive }\\ 
    \For{$j \in I$}{
        $(\beta_{G_j}, \betagej)$ = coordinate\_updates(\ ) \textcolor{blue}{//coordinate-wise solutions}
    }
 }
 \caption{cyclic\_coordinate\_updates(\textit{set of indices} $I$)}
\end{algorithm}
\vskip0.1in
\end{footnotesize}

\vskip0.1in
\begin{footnotesize}
\begin{algorithm}[H]
\SetAlgoLined
\For{i\_outer = 1, ... , max\_iter}{
    check\_duality\_gap($\beta$)\\
    d = compute\_d\_i(\ ) \\
    \eIf {i\_outer == 1}{
        working\_set = \{$j: \ (\beta_{G_j} \ne 0) \text{ or } (\betagej \ne 0)$\}\ \textcolor{blue}{ //initialization}\\
        \If {length(working\_set) == 0}{
            working\_set = \{j: smallest d[j] up to working\_set\_init\_size\}
        }
        working\_set\_size\ = length(working\_set)
    }{
        d[working\_set] = - Inf
        \textcolor{blue}{ //to make sure WS increases monotonically}
        \\
        working\_set\_size = min(2$\cdot$working\_set\_size, p) \textcolor{blue}{ //doubling the size of the WS at each iteration}\\
        working\_set = \{j: smallest d[j] up to working\_set\_size\}
        \textcolor{blue}{ //scoring the coef. by the likelihood of being non-zero}
    }
    \vskip0.1in
    cyclic\_coordinate\_updates(working\_set)\ \textcolor{blue}{ //\textbf{Algorithm 1} or optimized \textbf{Algorithm 3}} 
}
 \caption{coordinate\_descent\_with\_working\_sets(\ )}
\end{algorithm}
\end{footnotesize}
\vskip0.1in

\subsubsection{Active set and adaptive max difference strategies} \label{section_max_diff}
In Algorithm 1 the dual feasible point required to determine the dual gap is calculated according to the naive projection method proposed in section \ref{section_naive}. 
However, computing the naive projection is expensive, since it requires performing a matrix by vector product with a vector of length $n$ (sample size) and a matrix of size $n \times \text{length(working\_set)}$.
The idea of an active set and adaptive max difference strategies is to reduce the number of times we have to evaluate check\_duality\_gap() function to ensure convergence. 

\textbf{Adaptive max difference strategy:}
In any iterative optimization algorithm, including coordinate descent, the convergence (stopping) criterion plays an important role. One of the most commonly used stopping criteria is based on the change in estimates between two consecutive iterations $t-1$ and $t$ (\ref{stopping_1}). It is implemented in the glmnet package [\cite{glmnet}], for example. The idea is that if the estimated coefficients do not change much from iteration to iteration it is likely the  optimal solution has been reached:
\begin{align} \label{stopping_1}
    \max_{i \in \{1..p\}} \big| \beta_i^{(t)} - \beta_i^{(t-1)}\big|_2^2\  \norm{X_i}_2^2 < \epsilon.
\end{align}
However, in contrast to the duality gap stopping criterion, such heuristic rules do not offer control over suboptimality and it is generally not clear what value of $\epsilon$ is sufficiently small.

Although the heuristic convergence criterion based on the maximum absolute difference between consecutive estimates (\ref{stopping_1}) does not control the suboptimality gap $P(\beta^{(t)}) - P(\hat{\beta})$,  it is very fast to compute, since $\norm{X_i}_2^2$ is pre-computed or normalized to be one.
To reduce the number of times we check the convergence based on the duality gap criterion in Algorithm 1, we propose to use the criterion (\ref{stopping_1}) as a proxy convergence criterion (Algorithm 3). 

The proposed procedure is as follows: we initialize the tolerance for the max difference criterion as the tolerance we set for the duality gap convergence. We proceed by fitting the coordinate descent algorithm until we meet the max difference convergence criterion and then check the duality gap for the final convergence. If the duality gap criterion is not met, we decrease the max difference tolerance by a factor of 10 and proceed again. As a result, instead of checking the duality gap after each cycle of the coordinate descent algorithm, we wait until the proxy convergence criterion (which is very cheap to check) is met, and then check the duality gap criterion. By adaptively reducing the tolerance value of the proxy criterion we allow more coordinate descent cycles if needed, but carefully control the adaptive convergence to make as few checks as possible. 

\textbf{Active set strategy:}
The active set (AS) is a heuristic proposed by the authors of the glmnet package. The active set strategy tracks predictors updated during the first coordinate descent cycle and proceeds by fitting the coordinate descent algorithm only on those variables.  
The active set strategy combines naturally with our proposed adaptive max difference procedure since we also calculate the differences of the estimates on the consecutive iterations for our proxy convergence criterion. We believe that the max difference strategy we proposed in combination with the active sets can accelerate the state of art methods for fitting the standard Lasso as well. We summarize the main steps for both strategies in Algorithm 3, which is an optimized version of Algorithm 1.

To recapitulate, Algorithm 1 is the vanilla block coordinate descent algorithm, where each block in $\{1,..., p\}$, comprised of a main effect and its corresponding interaction, is updated until convergence without the use of any screening rule or heuristic. Algorithm 2 incorporates the Gap SAFE screening rules to the block coordinate descent algorithm in combination with the working set strategy. In the outer loop the variables in the working set are determined and in the inner loop the basic block coordinate descent Algorithm 1 is applied to the reduced set of variables. This achieves a lower run time. Algorithm 3 leverages additional heuristics to further accelerate the convergence on working sets by using active sets and the fast maximum absolute difference convergence criterion. Note that all the proposed improvements guarantee convergence to a global optimal solution since the final convergence criterion ensures that the duality gap for the full problem (including all the estimated variables) is within a given tolerance. 

\vskip0.1in
\begin{footnotesize}
\begin{algorithm}[H]
\SetAlgoLined
 \For{i\_inner = 1, ... , max\_iter}{
    check\_duality\_gap($\beta$)\\
    \eIf{i\_inner == 1}{max\_diff\_toll = tol \textcolor{blue}{ //initializing heuristic convergence criterion}
    }{
    max\_diff\_toll = max\_diff\_toll/10 \textcolor{blue}{ //adaptively refine the criterion}
    }
    \While{t $<$ max\_iter}{
    max\_diff = 0\\
    \For{$j \in I$}{
        $(\beta_{G_j}, \betagej)$ = coordinate\_descent\_update(\ ) \\
        current\_diff = max\Big(
        $\norm{\beta_{G_j}^{(t)} - \beta_{G_j}^{(t-1)}}^2\norm{G_j}^2,\norm{\betagej^{(t)} - \betagej^{(t-1)}}^2\norm{G_j\times E}^2 $\Big) \textcolor{blue}{ //max difference heuristic convergence criterion}\\
        max\_diff = max(current\_diff, max\_diff)\\
        \If{current\_diff $>$ 0}{active\_set[j] = TRUE \textcolor{blue}{ //active set = coefficients that got updated}}
    }
    \lIf{max\_diff $<$ max\_diff\_tol}{
    break \DontPrintSemicolon}
   
     \While{t $<$ max\_iter}{
     max\_diff = 0\\
     \For{$j \in$ active\_set }{
     \textcolor{blue}{ //active set iteration}\\
      $(\beta_{G_j}, \betagej)$ = coordinate\_descent\_update(\ )\\
        current\_diff = max\Big($\norm{\beta_{G_j}^{(t)} - \beta_{G_j}^{(t-1)}}^2\norm{G_j}^2,\norm{\betagej^{(t)} - \betagej^{(t-1)}}^2\norm{G_j\times E}^2 $\Big)\\
        max\_diff = max(current\_diff, max\_diff)
     }
     \lIf{max\_diff $<$ max\_diff\_tol}{
     break \DontPrintSemicolon
}}}}

 \caption{cyclic\_coordinate\_updates\_optimized(\textit{set of indices} $I$)}
\end{algorithm}
\end{footnotesize}
\vskip0.1in

\subsection{Experiments}
\begin{figure}[ht!]
\centering
\includegraphics[width=1\textwidth]{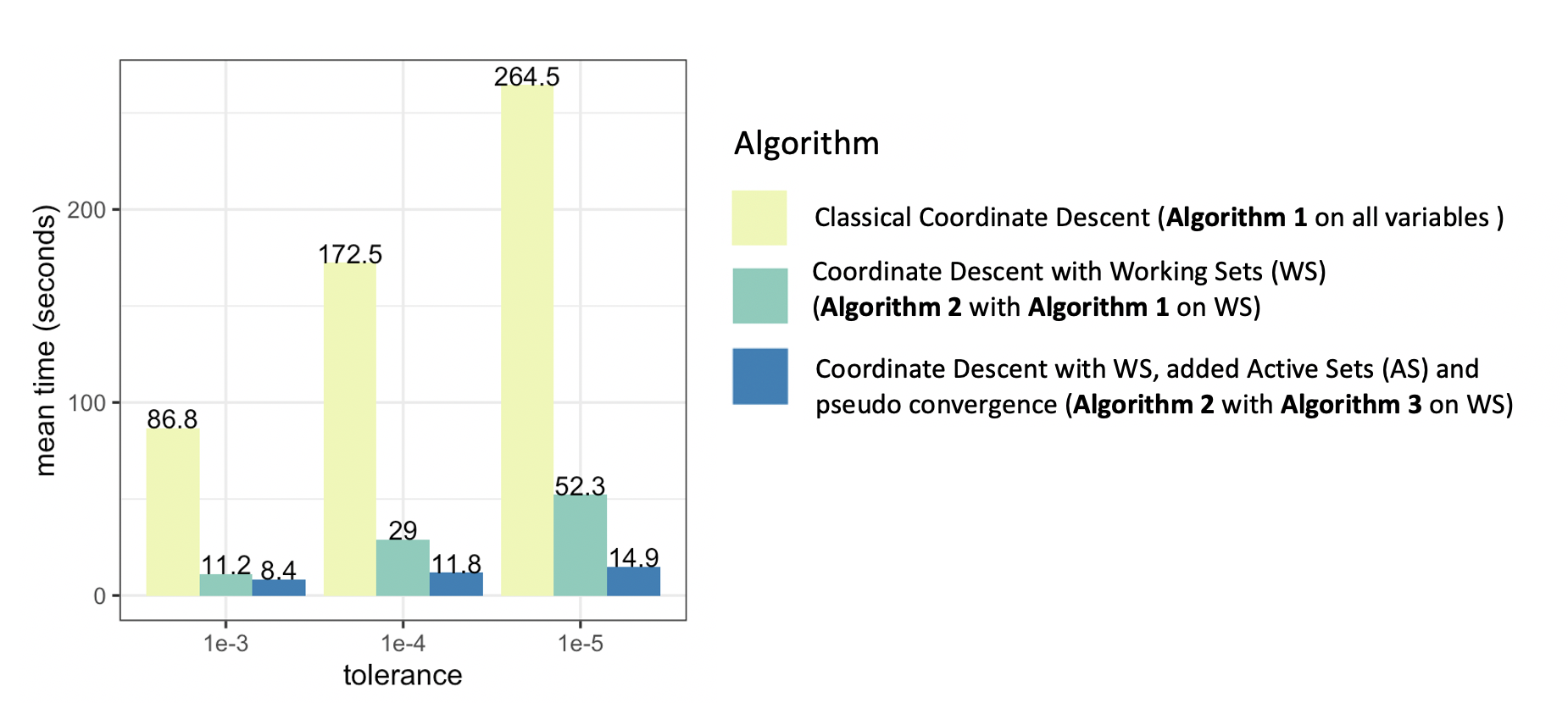}
\caption{Time comparison of proposed algorithms: mean runtime over 100 replicates on the y-axis, dual gap tolerance on the x-axis.}
\label{time_comparison}
\end{figure}

We conducted a series of experiments to evaluate efficiency of our algorithm overall and with respect to the screening rules.\\
\textit{Runtime:}
 In the first experiment we compared the runtime of the basic coordinate descent algorithm (Algorithm 1) and the various speedup strategies we described in the  sections above. We adopted the working set strategy for the \texttt{gesso} problem (Algorithm 2), we  proposed the max difference strategy, and we added the active set strategy in combination to the max difference approach (Algorithm 3). The size of the dataset we used for the experiment was $n = 200,\  p = \num{10000}\ (\num{20001}$ predictors in total, $p$ main effects and $p$ interaction terms, one environmental variable). We simulated 15 non-zero main effects and 10 non-zero interaction terms. We ran each algorithm 100 times and reported the mean execution time in Figure \ref{time_comparison}. Figure \ref{time_comparison} demonstrates that proposed speedup strategies result in a major run-time acceleration. Therefore, only the fastest algorithm containing all the proposed improvements (coordinate descent on WS with adding AS, and adaptive pseudo convergence) is implemented in the package and is used for the downstream experiments.
 
\begin{figure}[ht!]
\centering
\includegraphics[width=0.5\textwidth]{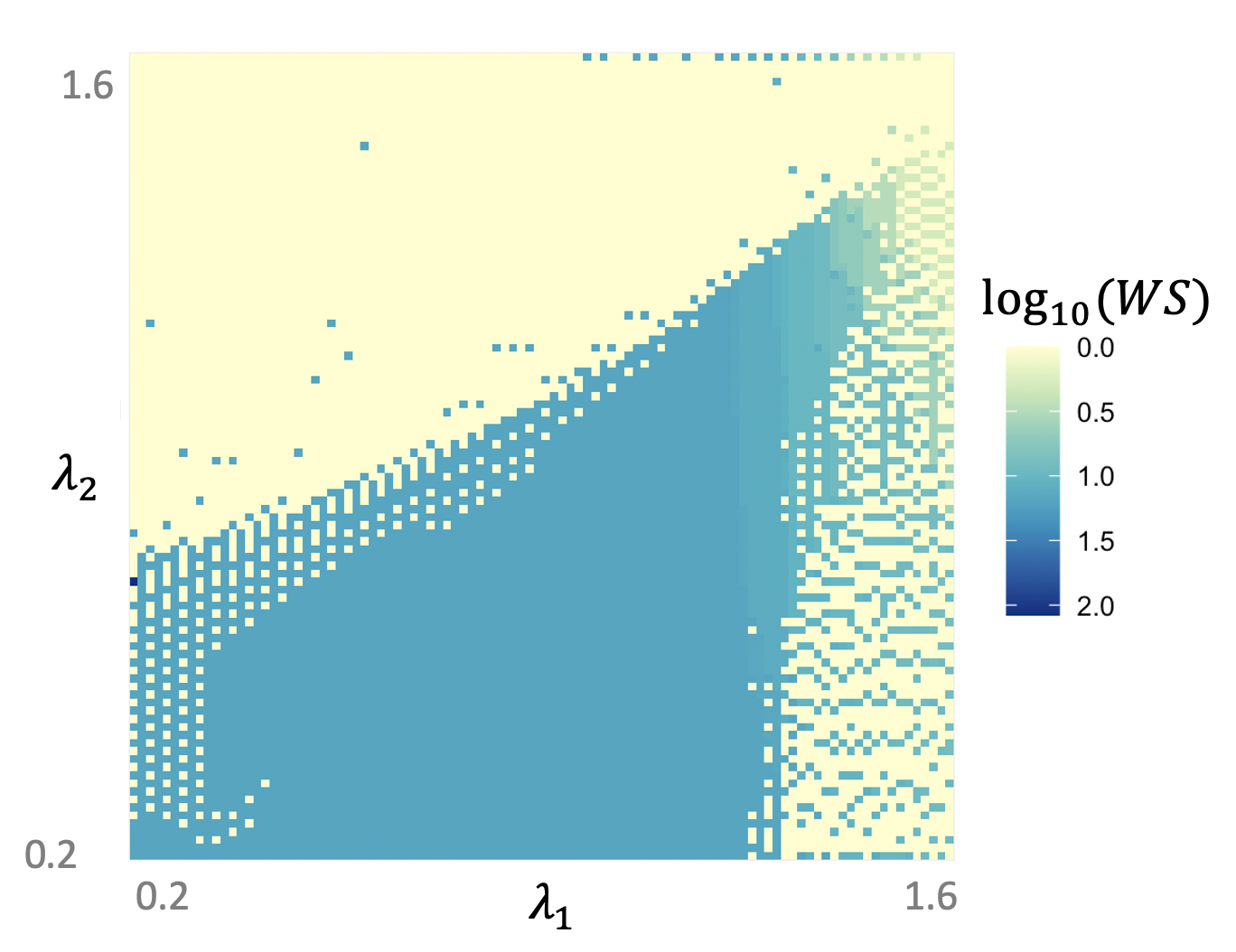}
\caption{Log working set size ($\log_{10} (WS)$) for all lambda pairs.}
\label{WS_size}
\end{figure}

\textit{Working sets:}
In the next experiment we evaluated the screening ability of our screening rules in combination with the 
working sets (WS) speedup. 
We simulated a dataset with $n=\num{1000}$ and $p = \num{100000}$, with non-zero main effects and 10 non-zero interaction terms. 
Figure \ref{WS_size} shows log ten size of the working set for all pairs $(\lambda_1, \lambda_2)$ of the tuning parameters. Maximum size of the working set is 120 (out of $\num{100000}$ predictors). For most pairs only a small number of variables (from 0 to 15) is needed to find the solution. 


\section{Simulations}

\begin{table}[h]
\begin{footnotesize}
  \caption{Hierarchical G$\times$E Models}
  \centering 
  \begin{threeparttable}
    \begin{tabular}{cll}
    Name  &  Model penalty & \makecell{Comments (convexity, assumptions,\\ hyper-parameters)}\\
     \midrule\midrule
gesso  & $ \lambda_1  \sum_{i=1}^p   \norm{(\beta_{G_i}, \beta_{G_i \times E})}_{\infty}   +  \lambda_2 \norm{\beta_{G \times E}}_1$ &   \makecell{convex, two tuning parameters,\\
$\beta_E$ is not penalized}\\ \cmidrule(l  r ){1-3}
glinternet & $\begin{aligned}
& \lambda \Big(|\beta_E^{(0)}| + \sum_{i=1}^p |\betagi^{(0)}| + 
\sum_{i=1}^p \norm{ (\beta_E^{(j)}, \betagj^{(E)}, \betagej)}_2 \Big), \\
&\text{where  } \betagj = \betagj ^{(0)} + \betagj ^{(E)}, \ \
\beta_E = \beta_E^{(0)} + \beta_E^{(1)} + \dots + \beta_E^{(p)}
\end{aligned}
$   &  \makecell{convex, single tuning parameter,\\
$\beta_E$ is penalized}\\ \cmidrule(l r ){1-3}
FAMILY &
 $\begin{aligned}
  (1 - \alpha) \lambda  \sum_{i=1}^p   \norm{(\beta_{G_i}, \beta_{G_i \times E_1}, .., \beta_{G_i \times E_q})}_{\infty}   +  \alpha \lambda \norm{\beta_{G \times E}}_1\\
  (1 - \alpha) \lambda  \sum_{i=1}^p   \norm{(\beta_{G_i}, \beta_{G_i \times E_1}, .., \beta_{G_i \times E_q})}_{2}   +  \alpha \lambda \norm{\beta_{G \times E}}_1
 \end{aligned}
 $ &  \makecell{$q > 1$, convex, two tuning parameters,\\
 $\beta_E$ is not penalized} \\ \cmidrule(l r ){1-3}
sail & $\begin{aligned}
& (1 - \alpha) \lambda \big( |\beta_E| + \sum_{i=1}^{p} |\betagj| \big) + \alpha \lambda \sum_{i=1}^{p} |\gamma_j|,\\
&\text{where  } \betagej = \gamma_j \beta_E \betagj
\end{aligned}
$&  \makecell{non-convex, one tuning parameter $\lambda$,\\ $\alpha$ has to be set,\\
$\beta_E$ is penalized} \\
    \midrule\midrule
    \end{tabular}
\end{threeparttable}
\end{footnotesize}
  \end{table}

We performed a series of simulations to evaluate the selection performance of \texttt{gesso} and compare it to that of alternative models. As a baseline model for comparison, we used the standard Lasso model as implemented in the glmnet package [\cite{glmnet}]. Among the models that impose hierarchical interactions, we chose the glinternet [\cite{glinternet}] model for comparison (Table 1), as it is implemented in an R package and can handle the G$\times$E case with a single environmental variable (by specifying the parameter \textit{interactionCandidates}), which is the focus of this paper. We also considered the FAMILY [\cite{family}] and sail [\cite{sail}] models and their respective packages for comparison. FAMILY implements a series of overlapped group lasso models (Table 1), but the package is designed for fitting
strictly more than one environmental variable $E$. We nevertheless modified the source code to handle the single $E$ case but were unable to achieve a satisfactory performance compared to other methods.
We omit FAMILY from the results.

The sail method ensures hierarchy via a reparametrization of the interaction coefficient that results in a non-convex objective formulation (Table 1). We observed that it performs similarly to the other examined methods in the cases where $p \approx n$, but for the high-dimensional setting we considered, the sensitivity for selecting interaction terms was poor and the execution time was comparably longer. In addition, sail only tunes the main penalty parameter, but not the relative weight of the penalties (parameter $\alpha$). In practice, the optimal relative weight is highly dependent on the data, and the default value equal to 0.5 yields poor selection in most cases. Because the weight parameter controls the relative importance of main effects and interactions, it plays a critical role in the selection of interaction performance. This is unlike the weight penalty parameter in elastic net, where relative importance varies between penalties on the same set of coefficients.

The hierarchical Lasso for pairwise interactions is implemented in the hierNet [\cite{bien}] R package. However, it can only handle a limited number of predictors and cannot be applied to the G$\times$E case.

\subsection*{Simulation settings}
We simulated data with $n=100$ subjects, $p = \num{2500}$ SNPs, and a single binary environmental variable $E$, for a total of $\num{5001}$ predictors. We set the number of non-zero main effects, $p_G$ out of $p$ SNPs to 10 and  the number of non-zero interaction, $p_{G\times E}$,  out of the $p$ interaction terms to 5. We simulated all non-zero main effects $\betag$ to have the same absolute value with randomly chosen signs and similarly for the interaction terms $\betage$. 

We explored three simulation modes for the true coefficients that we call \textit{strong\_hierarchical}, \textit{hierarchical}, and \textit{anti\_hierarchical}. 
In the strong hierarchical mode the hierarchical structure is maintained ($\beta_{G_i} = 0  \Longrightarrow \beta_{G_i \times E} = 0$) and also $|\beta_{G_i}| \ge |\beta_{G_i \times E}|$. In the hierarchical mode the  hierarchical structure is maintained, but we set $|\beta_{G_i}| \le |\beta_{G_i \times E}|$, with an attempt to violate the \texttt{gesso} assumption on the effect sizes. In the anti-hierarchical mode the hierarchical structure is violated ($\beta_{G_i \times E} \ne 0  \Longrightarrow \beta_{G_i} = 0$). We simulated a single binary environmental factor with  prevalence equal to 0.3. We set $\betag = 3$ and $\betage = 1.5$ for all of the modes except for the hierarchical mode where we set $\betag = 0.75$ and $\betage = 1.5$. We set the noise variable such that an interaction SNR (signal to noise ratio, defined as a ratio of the interaction signal to noise) is around 2.

We generated independent training, validation, and test sets under the same settings and report the model performance metrics on the test set. We run 200 replicates of the simulation for each parameter settings.
\begin{figure}[h!]
\includegraphics[width=1\textwidth]{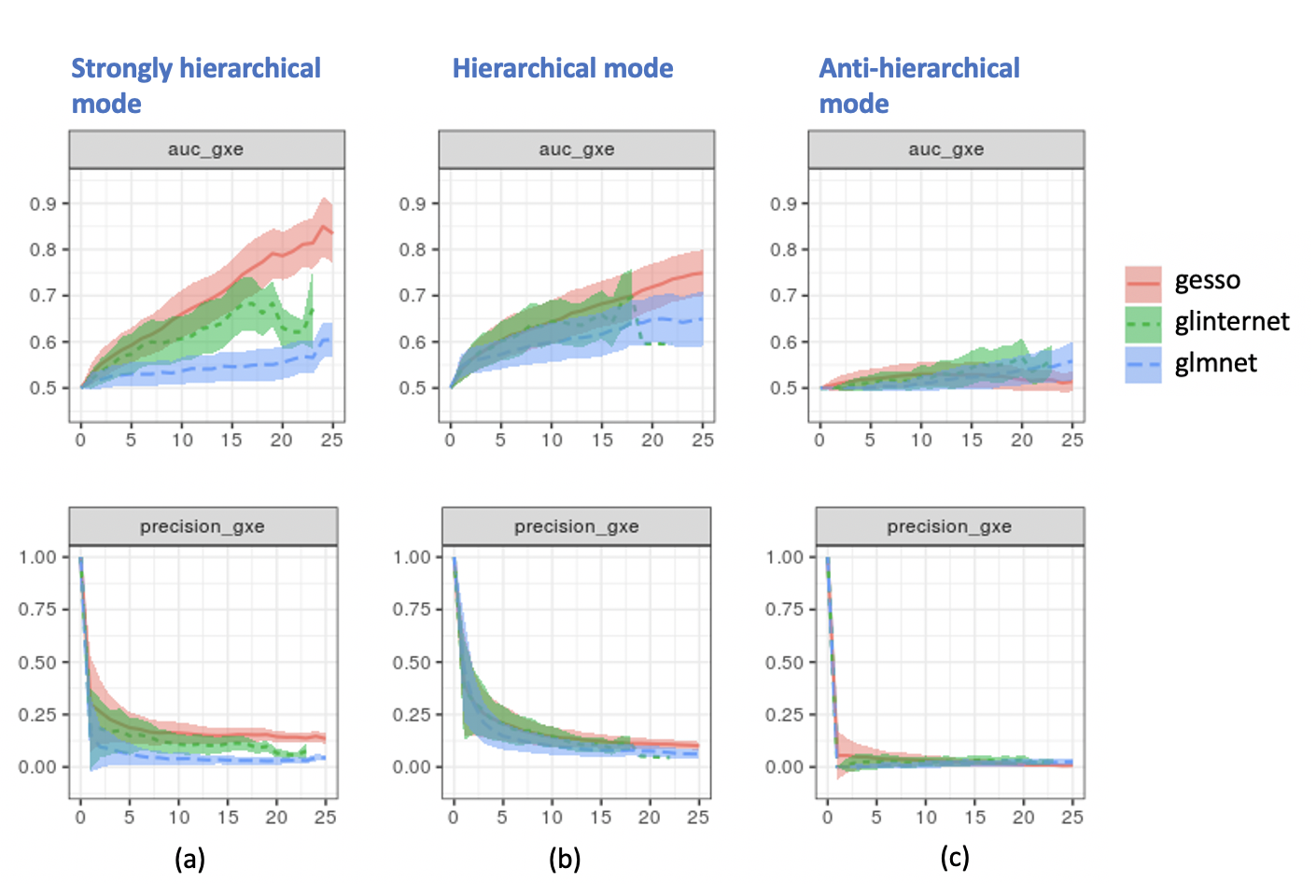}
\caption{Model performance (top row: AUC for G$\times$E selection, bottom row: precision for G$\times$E selection) as a function of the number of interactions discovered. p=2500, n=100, $p_G$=10, $p_{G\times E}$=5.}
\label{sim}
\end{figure}
\subsection*{Results}
We obtained solutions paths across a two-dimensional grid of tuning parameter values and computed the precision and area under the curve (AUC) for the detection of interactions as a function of the number of interactions discovered (Figure \ref{sim}).

In the strong hierarchical mode Figure \ref{sim}(a) the models imposing a hierarchical structure (\texttt{gesso} and glinternet) outperform the Lasso model. \texttt{gesso} performs better selection compared to glinternet, in terms of both AUC and precision.  

We considered the hierarchical mode to evaluate the selection performance of \texttt{gesso} when $|\betagei| \le |\betagi|$ does not hold. Because \texttt{gesso} does not make this stringent assumption, (only the laxer $|\betagei| \le \betagi^+ + \betagi^-$) we expect it to be robust to its violation.
In the hierarchical mode Figure \ref{sim}(b) glinternet and \texttt{gesso} still outperform the Lasso and \texttt{gesso} model performs on par with glinternet.

In the anti-hierarchical mode Figure \ref{sim}(c) all three models perform similarly. Importantly, even though the hierarchical assumptions are violated, hierarchical models still do not lose to the Lasso model in terms of selection performance. 


\section{Real data example: Epigenetic clock}
We analyzed the GSE40279 methylation dataset [\cite{geo_dataset}], that contains $\num{450000}$ CpG markers on autosomal chromosomes from whole blood for 656 subjects (Illumina Infinium 450k Human Methylation Beadchip), age, and gender. \cite{geo_dataset} reported that the methylome of men appeared to age approximately 4 percent faster than that of women. Epidemiological data also indicates that females live longer than males but the reasons for gender discrepancies are still unknown. 
Our goal in this analysis was to identify methylation by sex interactions that may associate with age. 

We preselected $\num{100000}$ of the most variable probes for our analysis, leaving us with $\num{100000}$ CpG probe main effects and $\num{100000}$  methylation probes by sex interaction terms, for a total of $\num{200001}$ predictors. Age is our dependent/outcome variable.  We ran our implementation of the \texttt{gesso} model over one hundred 5-fold cross-validation assignments and calculated the selection rate for each of the predictors as the number of times a predictor is selected by the model across the number of runs divided by the number of runs (one hundred in our case). We choose the hyper-parameters based on the minimum cross-validation loss.
We performed the same procedure for the standard Lasso model using the glmnet package.

\subsection*{Results}
\begin{table*}[h]
\begin{footnotesize}
\caption{Top interacting with sex CpG probes by method. }
\label{table0}
\begin{center}
\begin{tabular}{@{}llcc@{}}
\hline
CpG probe & Associated Gene & gesso & glmnet \\
\hline
\textbf{cg00091483}*    & PEBP1 & \textbf{1}  & \textbf{12}    \\
\textbf{cg12015310}    & MEIS2  & \textbf{2}  & \textbf{6}   \\
cg08327269    & NEUROD1  & 3  & 56    \\
cg103755456   & ICAM1  & 4  & 163  \\
\hline
\textbf{cg14345497}    & HOXB4  & \textbf{8}  & \textbf{1} \\
cg09365557    & TBX2 & $>$50,000  & 2   \\
cg27652200    & PSMB9/TAP1  & 109  & 3  \\
\textbf{cg19009405}    & HNRNPUL2  & \textbf{14}  & \textbf{4}  \\
\hline
\end{tabular}
\end{center}
*Probes identified by both methods (ranked among the top 20 probes by the
other method) are highlighted in bold.
\end{footnotesize}
\end{table*}
The average run-time for our implementation of \texttt{gesso} across 100 runs of cross-validation was only 4 minutes for the grid of 30 values for both of our tuning parameters. The average run-time for the glmnet package for the same grid for one tuning parameter was 1.5 minutes. 

We attempted to use the glinternet package for the analysis but it exited with an error due to the large data size. We then downloaded the source code and changed the memory handling code since it was allocating memory for the  G$\times$G before subselecting it to the G$\times$E case. However, even with the updated function glinternet would not complete the analysis within 24 hours. We hypothesize that, due to specifics of the current implementation, the glinternet package is efficient for the symmetric G$\times$G format, but not for the reduced G$\times$E format with large G. We also tried to run the FAMILY package on our methylation dataset after adapting the source code to generalize the function to the G$\times$E case with a single E variable. However, the program exited with an out of memory error. The sail package also did not finish within 24 hours. 
To conclude, the existing packages are not able to handle large datasets (here $\num{200000}$ variables) for the G$\times$E analysis.  

Table \ref{table0} presents the four top-ranking CpG probes for \texttt{gesso} and standard Lasso that interact with sex in their effects on age. Ranks were calculated by ordering interaction selection rates and assigning rank one to the highest rate value (most frequently selected variable), and so on.

Genes associated with the probes selected by the \texttt{gesso} were linked to aging processes and cell senescence in multiple publications and databases, and are suggested biomarkers for age-related diseases [\cite{gwas_ts}, \cite{p21_1},  \cite{icam2_1}]. 
For example, PEBP1 gene (linked to the top selected probe) is involved in the aging process and negative regulation of the MAPK pathway [\cite{peb-mapk}]. The MAPK and SAPK/JNK signaling networks promote senescence (in vitro) and aging (in vivo, animal models and human cohorts) in response to oxidative stress and inflammation [\cite{MAPK_aging}]. The Rat Genome Database (RGD) indicates that the PEBP1 gene is implicated in prostate and ovarian cancers [\cite{ovarian}], indicating some sex-specificity. 
The RGD database reports the PEBP1 gene as a biomarker of Alzheimer's disease. 

Among the top four probes based on the standard Lasso analysis, three probes were linked to regulatory genes. For example, TBX2 (linked to cg09365557) encodes  transcription factor that, when up-regulated, inhibits CDKN1A (p21), the gene regulated by the NEUROD1 gene (cg08327269 probe) discovered by \texttt{gesso} [Gene Cards Database]. CDKN1A is necessary for tissue senescence, and when compromised, leaves the tissue vulnerable to tumor-promoting signals. Probes cg09365557 (identified by standard Lasso) and cg08327269 (identified by \texttt{gesso}) could both be uncovering the same biological process involving CDKN1A.

\section{Discussion}
We introduced a selection method for G$\times$E interactions with the hierarchical main-effect-before-interaction property.
We showed that existing packages for the hierarchical selection of interactions cannot handle the large number of predictors typical in studies with high-dimensional omic data. When considered separately, and not as a sub-case of a G$\times$G analysis, the G$\times$E case can be solved much more efficiently. Our proposed block coordinate descent algorithm is scalable to large numbers of predictors because of the custom screening rules we developed. We also showed in simulations that our model outperforms other hierarchical models. The implementation of our method is available in our R package \texttt{gesso} that can be downloaded from CRAN \url{https://CRAN.R-project.org/package=gesso}. Our algorithm can be extended to generalized linear models via iteratively reweighted least-squares.

The model can be generalized to include more than one environmental variable $E$. However, for the BCD algorithm to be efficient, blocks have to be relatively small. When including multiple $E$ variables, the size of the coordinate descent blocks grows linearly and contains main effect and all its corresponding interactions with the environmental variables $(\betagj, \beta_{G_j \times E_1}, ...  \beta_{G_j \times E_q})$. The bottleneck here is to efficiently solve the system of equations resulting from the stationarity conditions, the size of which grows exponentially with the number of $E$ variables. For more than a handful of $E$ variables other algorithms, like ADMM, could become more efficient.

Apart from computational efficiency of the algorithm, memory considerations are critical for large-scale analyses.  The \texttt{gesso} package allows users to analyze large genome-wide datasets that do not fit in RAM using the file-backed bigmemory [\cite{bigmemory}]  format. However, this involves  more time-consuming data transfers between RAM and an the external memory source.  To more efficiently handle datasets that exceed available RAM,  the screening rules could be exploited for efficient batch processing [\cite{basil}].

\bigskip
\begin{center}
{\large\bf ACKNOWLEDGEMENTS}
\end{center}
We sincerely thank Jacob Bien, Paul Marjoram, and David Conti for their helpful comments on this work.

\bigskip
\begin{center}
{\large\bf FUNDING}
\end{center}
Research reported in this paper was supported by NCI of the National Institutes of Health under award number P01CA196569, FIGI RO1 supported by NCI (R01CA201407), and T32 supported by NIEHS (T32ES013678).

\begin{center}
{\large\bf APPENDIX}
\end{center}
\section*{Appendix A: Proof of equivalence of relaxed and unconstrained models} \label{apx:equivalence}
We want to prove that models (\ref{relaxation}) and (\ref{unconstrained}) are equivalent: 
$$
 \min_{\beta_0, \beta_G^{\pm}, \beta_E, \beta_{G\times E}}
q(\beta_0, \beta_G^{\pm}, \beta_E, \beta_{G\times E}) + \lambda_1 ( \beta_{G_i}^+ + \beta_{G_i}^- ) + \lambda_2 \norm{\beta_{G\times E}}_1,
$$
$$
\text{subject to: }
|\beta_{G_i \times E}| \le \beta_{G_i}^+ + \beta_{G_i}^-, \ \beta_{G_i}^{\pm} \ge 0 \text{ for } i = 1, ..., p.
$$
and 
$$
\min_{\beta_0, \beta_G, \beta_E, \beta_{G\times E}}
q(\beta_0, \beta_G, \beta_E, \beta_{G\times E}) + \lambda_1 \sum_{i=1}^p \max \big\{ |\beta_{G_i}|, |\beta_{G_i \times E}|  \big\}  + \lambda_2 \norm{\beta_{G \times E}}_1.
$$
\textbf{Proof}: Recall that
$\beta_{G_i} =  \beta_{G_i}^+ - \beta_{G_i}^-$ and $\beta_{G_i}^{\pm} \ge 0$, then $\beta_{G_i}^- =  \beta_{G_i}^+ - \beta_{G_i}$ and \\ $\beta_{G_i}^+ \ge \beta_{G_i}$. $ \beta_{G_i}^+ + \beta_{G_i}^- = 2 \beta_{G_i}^+ -  \beta_{G_i}$ and since $|\beta_{G_i \times E}| \le \beta_{G_i}^+ + \beta_{G_i}^-$, we have $\beta_{G_i}^+  \ge \frac{|\beta_{G_i \times E}| + \beta_{G_i}}{2}$.
Since $\beta_{G_i}^+ \ge 0$, $\beta_{G_i}^+ \ge \beta_{G_i}$, and $\beta_{G_i}^+  \ge \frac{|\beta_{G_i \times E}| + \beta_{G_i}}{2}$, we can write $\beta_{G_i}^+ \ge \max \Big \{ [\beta_{G_i}]_+ , \ \frac{|\beta_{G_i \times E}| + \beta_{G_i}}{2} \Big \}$, where $[\beta_{G_i}]_+ = \max \{\beta_{G_i},\ 0 \}$. Then we have that $\beta_{G_i}^+ + \beta_{G_i}^- = \\
=2 \beta_{G_i}^+ -  \beta_{G_i} \ge \max \Big \{2[ \beta_{G_i}]_+ - \beta_{G_i} ,\ |\beta_{G_i \times E}|  \Big \}$. Finally, we note that $2 [\beta_{G_i}]_+ -  \beta_{G_i}  = |\beta_{G_i}|$. We have $\beta_{G_i}^+ + \beta_{G_i}^- \ge \max \Big \{|\beta_{G_i}|,\ |\beta_{G_i \times E}|  \Big \}$ and since we are solving a minimization problem we can substitute $( \beta_{G_i}^+ + \beta_{G_i}^- )$ in model (\ref{relaxation}) to its minimum value $ \max \Big \{|\beta_{G_i}|,\ |\beta_{G_i \times E}|  \Big \}$ which finalize our transformation from model (\ref{relaxation}) to (\ref{unconstrained}) and concludes the proof of equivalence. 

\section*{Appendix B: Solution to the optimization problem (\ref{safe_new})}
Consider an optimization problem:
\begin{maxi*}|l|
{x, \delta}{D(\nu_0)}
{}{}
\addConstraint{|\nu_0^T G\times E| \preccurlyeq\lambda_2 + \delta}
\addConstraint{|\nu_0^T G| \preccurlyeq \lambda_1 - \delta}
\addConstraint{\delta  \succcurlyeq 0}
\addConstraint{\text{where } \nu_0 = x \nu_{\mathrm{res}}(\beta)}
\addConstraint{D(\nu) = \frac{n}{2} \Big( \norm{\frac{Y}{n}}^2 -\norm{\frac{Y}{n} - \nu}^2
\Big)}.
\end{maxi*}
Lets denote $Y/n$ as $y$ and $\nu_{\mathrm{res}}(\beta)$ as $\nu$.
We have
\begin{mini}|l|
{x, \delta}{\norm{y - x\nu}^2}
{\label{3.6}}{}
\addConstraint{|x\nu^T G_j\times E| \le \lambda_2 + \delta_j}
\addConstraint{|x\nu^T G_j| \le \lambda_1 - \delta_j}
\addConstraint{\delta_j \ge 0, \text{ for } j = 1..p.}
\end{mini}
Solution:
$$
\frac{\partial D(\nu)}{\partial x} = (y - x\nu)^T\nu = 0,\ \ 
y^T \nu -  x \norm{\nu}_2^2 = 0, \ \
\hat{x} =  \frac{y^T \nu}{\norm{\nu}_2^2}.
$$
Lets denote $|\nu^T (G \times E)_j|$ as $B_j$ and $|\nu^T G_j|$ as $A_j$. The feasible set for our optimization  problem (\ref{3.6}) is
$$
\begin{cases}
|x| B_j \le \lambda_2 + \delta_j\\
|x| A_j \le \lambda_1 - \delta_j, 
\end{cases} \implies
\begin{cases}
|x| \le \frac{\lambda_2 + \delta_j}{B_j}\\
|x| \le \frac{\lambda_1 - \delta_j}{A_j}
\end{cases}.
$$
\begin{figure}[ht!]
\centering
\includegraphics[width=0.5\textwidth]{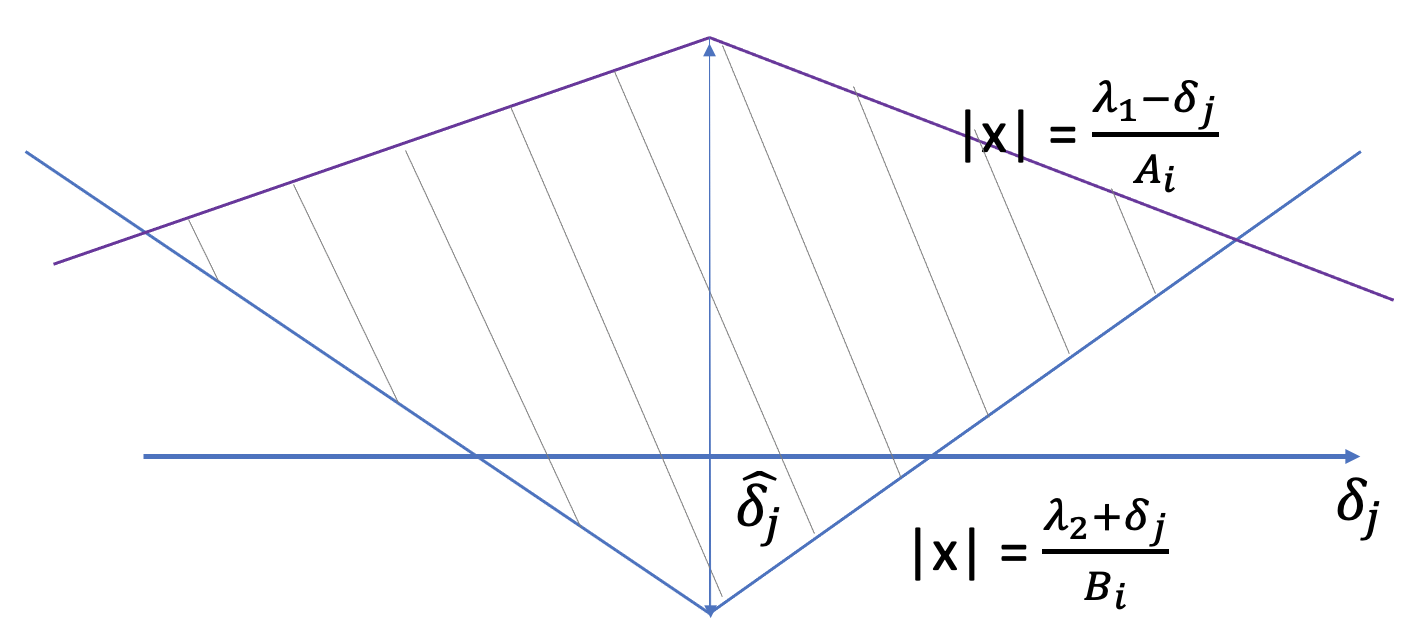}
\caption{Geometric solution to the  problem (\ref{3.6}). Optimal $\delta_j$ maximizes the range of possible $x$ values.}
\label{optimal_delta}
\end{figure}
Then optimal $\delta_j$ is the solution to $\frac{\lambda_2 + \delta_j}{B_j} = \frac{\lambda_1 - \delta_j}{A_j}$ because a solution to this equation maximizes the feasible set and hence provides more broad set to find an optimal $x$ (Figure \ref{optimal_delta}).
$$
\frac{\lambda_2 + \delta_j}{B_j} = \frac{\lambda_1 - \delta_j}{A_j} \implies
A_j (\lambda_2 + \delta_j) = B_j (\lambda_1 - \delta_j) \implies
\delta_j = \frac{B_j \lambda_1 - A_j \lambda_2}{B_j + A_j}.
$$
We know that $\delta_j \in [0, \lambda_1]$,  $ \forall j$ (section 1 of the supplementary materials). If $\delta_j > \lambda_1$, then $\hat{\delta}_j = \lambda_1$, and if $\delta_j < 0$, then $\hat{\delta}_j = 0$. Therefore, $$
\hat{\delta}_j = \max(0, \min (\delta_j, \lambda_1)).
$$
We found $\hat{x} = \frac{y^T \nu}{\norm{\nu}_2^2}$ that minimises our objective function by the stationarity conditions. To be optimal it has to satisfy out feasibility conditions for any j:
$$
\begin{cases}
|x| \le \min_j \frac{\lambda_1 - \hat{\delta}_j}{A_j},\\
|x| \le \min_j \frac{\lambda_2 + \hat{\delta}_j}{B_j}.
\end{cases}
$$
Then feasible x is going to satisfy the inequality
$$
|x| \le \min \Big( \min_j \frac{\lambda_1 - \hat{\delta}_j}{A_j}, \min_j \frac{\lambda_2 + \hat{\delta}_j}{B_j} \Big).
$$
Let's denote the RHS of the inequality as M. If $|\hat{x}| \le M$, then optimal $x$ = $\hat{x}$ and if $|\hat{x}| > M$, then
optimal $x$ = sign$(\hat{x})M$.

\section*{Appendix C: Safe rules for \texttt{gesso}}
By the KKT conditions (\ref{beta_j_zero}) and (\ref{beta_gxej_zero}):
\begin{align*} 
\begin{cases} 
\Big|\hat{\nu} ^T G_i \Big| < \lambda_1 - \delta_i \\
\Big|\hat{\nu} ^T (G_i\times E) \Big| < \lambda_2 + \delta_i \implies \hat{\beta}_i = \hat{\beta}_{G_i \times E} = 0.\\
\delta_j \in [0, \lambda_1]
\end{cases}
\end{align*}
Then,
\begin{align}
\begin{cases} \label{safe_start2}
\max_{\nu \in B(c, r)}\Big|\nu^T G_i \Big| < \lambda_1 - \delta_i \\
\max_{\nu \in B(c, r)}\Big|\nu ^T (G_i\times E) \Big| < \lambda_2 + \delta_i \implies \hat{\beta}_i = \hat{\beta}_{G_i \times E} = 0\\
\delta_i \in [0, \lambda_1]
\end{cases}
\end{align}
$$\iff
\begin{cases}
r  \norm{G_i} + |G_i^T c| <  \lambda_1 - \delta_i\\
r  \norm{G_i\times E} + |(G_i\times E)^T c| <  \lambda_2 + \delta_i \implies \hat{\beta}_i = \hat{\beta}_{G_i \times E} = 0\\
\delta_i \in [0, \lambda_1]
\end{cases}
$$
Let's consider the last system of inequalities more closely:
$$
\begin{cases}
r  \norm{G_i} + |G_i^T c| <  \lambda_1 - \delta_i\\
r  \norm{G_i\times E} + |(G_i\times E)^T c| <  \lambda_2 + \delta_i \\
\delta_i \in [0, \lambda_1]
\end{cases}
\iff
\begin{cases}
\delta_i < \lambda_1 - r  \norm{G_i} - |G_i^T c|\\
\delta_i > r  \norm{G_i\times E} + |(G_i\times E)^T c| -  \lambda_2  \iff \\
\delta_i \in [0, \lambda_1]
\end{cases}
$$
$$
\begin{cases}
\delta_i < \lambda_1 - r  \norm{G_i} - |G_i^T c|\\
\delta_i > \max \big\{0, \ r  \norm{G_i\times E} + |(G_i\times E)^T c| -  \lambda_2\big\}
\end{cases}
$$
Then, 
$$
(\ref{safe_start2}) \iff
\exists \ \delta_i \text{ feasible } \iff
\max \big\{0, \ r  \norm{G_i\times E} + |(G_i\times E)^T c| -  \lambda_2\big\} < \lambda_1 - r  \norm{G_i} - |G_i^T c|,
$$
and, hence, the SAFE rules to discard $(\betagi, \beta_{G_i \times E})$ are:
\begin{align} \label{SAFE_g_gxe_A}
\max \big\{0, \ r  \norm{G_i\times E} + |(G_i\times E)^T c| -  \lambda_2\big\} < \lambda_1 - r  \norm{G_i} - |G_i^T c|\\ \implies \hat{\beta}_i = \hat{\beta}_{G_i \times E} = 0.\nonumber
\end{align}

\begin{center}
{\large\bf SUPPLEMENTARY MATERIALS}
\end{center}

\begin{description}
\item[Supplement to “A scalable hierarchical lasso for gene-environment interactions”:]
We include detailed derivations for the coordinate-wise solutions (section 1) and the dual formulation (section 2).

\end{description}

\end{document}